\documentclass[aps,pra,floatfix,superscriptaddress,showpacs,twocolumn,amssymb]{revtex4}

\usepackage{natbib}
\usepackage{amsmath}
\usepackage{mathrsfs}
\usepackage{graphicx}
\usepackage{dcolumn}
\usepackage{bm}
\usepackage{euscript}

\usepackage[dvips]{color}

\def\rm#1{{\textrm{#1}}}

\def\prn#1{{\left(#1\right)}}

\def\sbrk#1{{\left[#1\right]}}

\begin{document}
\pagenumbering{arabic}
\setcounter{footnote}{0}

\title{Cancellation of nonlinear Zeeman shifts with light shifts}

\author{K.~Jensen}
\email{kjensen@nbi.dk}
\affiliation{Niels Bohr Institute, University of Copenhagen, DK 2100, Denmark}
\affiliation{QUANTOP, Danish National Research Foundation Center for Quantum Optics}

\author{V.~M.~Acosta}
\affiliation{Department of Physics, University of California at
Berkeley, Berkeley, California 94720-7300}

\author{J.~M.~Higbie}
\affiliation{Department of Physics and Astronomy, Bucknell University, Lewisburg, Pennsylvania 17837}

\author{M.~P.~Ledbetter}
\affiliation{Department of Physics, University of California at
Berkeley, Berkeley, California 94720-7300}

\author{S.~M.~Rochester}
\affiliation{Department of Physics, University of California at
Berkeley, Berkeley, California 94720-7300}

\author{D. Budker}
 \email{budker@berkeley.edu}
 \affiliation{Department of Physics, University of California at
Berkeley, Berkeley, California 94720-7300}
 \affiliation{Nuclear Science Division, Lawrence
 Berkeley National Laboratory, Berkeley, California 94720}

\begin{abstract}
Nonlinear Zeeman (NLZ) shifts arising from magnetic-field mixing of the two hyperfine ground-states in alkali atoms lead to splitting of magnetic-resonance lines. This is a major source of sensitivity degradation and the so-called ``heading errors'' of alkali-vapor atomic magnetometers operating in the geophysical field range ($B\approx0.2-0.7$ G). Here, it is shown theoretically and experimentally that NLZ shifts can be effectively canceled by light shifts caused by a laser field of appropriate intensity, polarization and frequency, a technique that can be readily applied in practical situations.
\end{abstract}
\maketitle
\section{Introduction}
Alkali-vapor atomic magnetometers \cite{Bud07} operating in the geophysical range of magnetic fields, where the Zeeman effect is very close to linear, are nevertheless quite sensitive to the nonlinear corrections to the Zeeman effect (NLZ) which cause broadening and splitting of the magnetic-resonance lines, as well as line-shape asymmetries that depend on the orientation of the sensor with respect to the field \cite{Aco06,Sel2007}. Thus, NLZ is responsible for sensitivity degradation and systematic (heading) errors in atomic magnetometers \cite{Ale03,Roch08}.
Recently, collapse and revival of ground-state quantum beats associated with NLZ was studied theoretically \cite{Ale2005} and experimentally \cite{Sel2007}, and a scheme for mitigating the effects of NLZ in an atomic magnetometer based on double-modulated synchronous optical pumping was realized \cite{Sel2007}. Alternative approaches explored recently include the use of multi-quantum transitions and high-order atomic polarization moments \cite{Ale97,Oku2001,Aco2008,Yas2003}.
Here, we introduce an alternative technique where NLZ shifts are compensated by light shifts due to an additional light field of appropriate intensity, polarization and frequency, which is possible due to the identical tensor structure of the splitting caused by the two effects. 
The present technique is free from some shortcomings of
the alternative techniques such as complexity of implementation and/or degradation of the signal. Moreover, in contrast to the
double-modulated synchronous pumping technique of Ref.\ \cite{Sel2007}, the present approach works well when the ground-state polarization-decay rate approaches the NLZ frequency splitting (a common situation for practically important Rb and Cs magnetometers).

Compensation of NLZ with AC Stark shifts was recently investigated in the field of quantum information, theoretically as part of a quantum memory protocol \cite{Opa06} and experimentally for improving atomic spin-squeezing \cite{Fer08}.
The similarities of light shifts and Zeeman shifts was used in Refs.\ \cite{Par02,Ven07}, where magnetic fields were simulated by light fields. It has also been proposed to use light shifts to measure parity violation in Fr atoms \cite{Bou08}. In general, light shifts are important in precision measurements in atomic physics, for instance in atomic clock experiments, where one can use the famous ``magic wavelengths'' to eliminate light shifts, see for example Refs.\ \cite{Ye08,Flam08} and references therein.

\section{Theory}
For an alkali atom with nuclear spin $I$, total angular momentum $F=I\pm 1/2$ and projection $M_F$ of the total angular momentum on the direction of the magnetic field, the Zeeman energies of the magnetic sublevels are to second order in the field $B$ given by
\begin{equation}
\Delta E \approx \pm \frac{2}{2I+1}\mu_B M_F B \pm\frac{\prn{\mu_B
B}^2}{\Delta_{\rm{hfs}}}\sbrk{1-\frac{4M_F^2}{\prn{2I+1}^2}},
\end{equation}
where $\mu_B$ is the Bohr magneton, $\Delta_{\rm{hfs}}$ is the hyperfine interval, and we neglect small corrections proportional to the nuclear magneton and the anomalous magnetic moment of the electron. The first-order shift leads to precession of the atomic polarization around the magnetic field with Larmor frequency $\nu_L=  2\mu_B B/\sbrk{h\prn{2I+1}}$, while the second-order shift leads to splitting of the magnetic resonances. For the $F=2$ hyperfine manifold of $^{87}$Rb, there are three $\Delta M_F=2$ resonances with adjacent-frequency splittting 
\begin{equation}
\delta\approx \frac{\mu_B^2 B^2 }{h\Delta_{\rm{hfs}}} .\label{NLZ_delta} 
\end{equation}
For $^{87}$Rb atoms placed in a magnetic field of 0.5 G, we have $\nu_L=350$ kHz and $\delta=72$ Hz.

Consider now $^{87}$Rb atoms illuminated with linearly polarized light near-resonant with the D1 transition. Based on the calculations in Refs.\ \cite{Hap67b,Mat68} we can find analytical results for the AC Stark shifts of the energy levels in the $F=2$ ground-state manifold. For light polarized along the magnetic field, we find a change in the splitting of adjacent $\Delta M_F=2$ resonance frequencies of
\begin{equation}
    \Delta^{D1}_{\pi}=-\frac{\lambda^3 \epsilon_0 E_0^2 \Gamma_{D1} A_{P_{1/2}}}
    {16\pi^2 h\Delta\prn{\Delta-2 A_{P_{1/2}}}},	\label{ShiftD1pi}
\end{equation}
where $\epsilon_0$ is the permittivity of vacuum, $A_{P_{1/2}}$ is the $5^2 P_{1/2}$ excited-state hyperfine-structure coefficient measured in Hz, 
$\Gamma_{D1}$ is the natural linewidth of the $5^2 P_{1/2}$ excited state, $\lambda$ is the transition wavelength, $E_0$ is the electric field amplitude of the light and the light frequency detuning $\Delta$ is measured relative to the $F=2\rightarrow F'=1$ transition and assumed to be much larger than the Doppler width and the upper-state hyperfine interval. 
For light polarized transverse to the magnetic field, we find that the change in splitting of the $\Delta M_F=2$ resonances are of half the 
magnitude and of the opposite sign compared to the longitudinal case.
In general, if the L2 light is
linearly polarized at an angle $\theta$
to the magnetic field, the differential shifts of the $M_F$ sublevels
are proportional to $3\cos^2 \theta-1$, see, for example, Problem 2.11 in Ref.\ \cite{AMObookBudker}.
Since the splitting increases in response to light of transverse polarization and decreases in response to longitudinal polarization, NLZ can only be compensated on this transition using longitudinal polarization.
Using Eqs.\ (\ref{NLZ_delta}) and (\ref{ShiftD1pi}) we can calculate the D1 light power needed to cancel the nonlinear Zeeman effect, and we find
\begin{equation}
P=\frac{\mu_B^2 B^2 }{\Delta_{\rm{hfs}} } \frac{2 \pi^3 c d^2 \Delta \left( \Delta-2A_{P_{1/2}}\right)}{\lambda^3 \Gamma_{D1} A_{P_{1/2}}}. \label{PowerD1Analytical}
\end{equation}
Here we asssume that the atoms are kept in a cylindrical cell of diameter $d$, and that the average light intensity $I=c \epsilon_0 E_0^2/2=P/\left(\pi d^2/4\right)$¨, where $c$ is the speed of light in vacuum, determines the amount of compensation. This is the case for atoms in an antirelaxation-coated cell, where each atom samples the light field in the entire cell volume.
Similar calculations for light near-resonant with the D2 transition shows that the change in splitting is of the opposite sign compared to the D1 case, and that roughly 8-10 times more power is needed to cancel NLZ.

In addition to shifting the magnetic resonances, light also broadens them. One source of this broadening is absorption of light by the ground-state atoms. 
The broadening depends on the light power needed to compensate NLZ.
For large detunings the broadening at the compensation power will be independent of detuning, since the power needed to cancel NLZ goes as $\Delta^2$ and the absorption rate goes as $1/\Delta^2$. Based on the calculations in Refs. \cite{Hap67b,Mat68}, we find the broadening at the compensation power for D1 light with longitudinal polarization to be
\begin{equation}
    \Gamma^{D1}_{\pi}=\frac{\mu_B^2B^2\Gamma_{D1}}{2 h A_{P_{1/2}}\Delta_{\rm{hfs}}}
     .  \label{BroadAbsD1}
\end{equation}
For D2 light with transverse polarization we find that the broadening at the compensation power is around 20 times larger compared to the D1 case. It will therefore be preferable to use D1 light to cancel NLZ.

Another source of light broadening results from the nonuniform intensity profile of the light beam over the cross-sectional area of the cell. If, as in our case, the compensating light illuminates only a small portion of the cell, the atoms are subject to short periods of compensating light at random intervals, so that the phase of each atom's evolution undergoes a random walk. This leads to dephasing of the atomic evolution at a rate
$\Gamma_{\text{dephase}}=\phi^2 /\tau$,
where $\phi$ is the phase shift an atom experiences in each pulse of compensating light, and $\tau$ is the average time between pulses. In order that the light properly compensates the nonlinear Zeeman effect, we must have $\phi=\delta\tau$. The average time for an atom to cross the cell is on the order of $d/v$, where $v$ is the rms atomic velocity, and the probability that an atom passes through the light beam in one trip across the cell is on the order of $b/d$, where $b$ is the beam diameter, so $\tau\approx d^2/(bv)$. Thus 
\begin{equation}
    \Gamma_{\text{dephase}} \approx \tau\delta^2
    \approx\frac{d^2\mu_B^4B^4}{bv\hbar^2\Delta_{\rm{hfs}}^2}
    . \label{GammaDephase}
\end{equation}
By inserting typical experimental values ($B=0.397$ G, $d\approx2$ cm and $b\approx 1.5$ mm), we find $\Gamma^{D1}_{\pi}\approx 0.3$ Hz and $\Gamma_{\rm{dephase}} \approx 11$ Hz, and we see that dephasing dominates the broadening.
\begin{figure}
    \includegraphics[width=3 in]{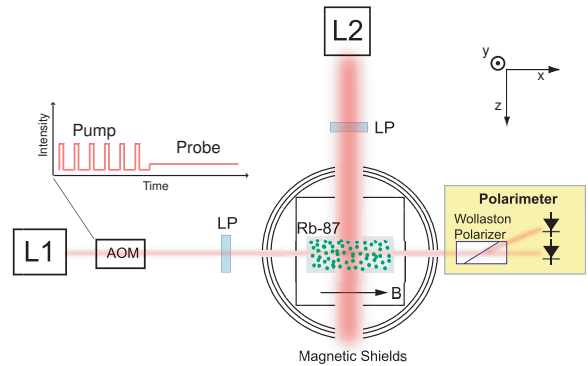}
    \caption{(Color online) Schematic of the experimental apparatus. AOM -- acousto-optical modulator; LP -- linear polarizers. A vapor cell with $^{87}$Rb is located within a multi-layer magnetic shield. Coils within the inner shield is used to apply the Earth-range magnetic field and to compensate gradients. The intensity of light from laser L1 is modulated as shown in the upper inset, so the same laser beam is used to optically pump and probe the atoms. Laser L2 is used to induce light shifts.}
    \label{Fig_Apparatus} 
\vspace{-.5cm}
\end{figure}

\section{Experimental setup, procedure and results}
\begin{figure}
    \includegraphics[width=3 in]{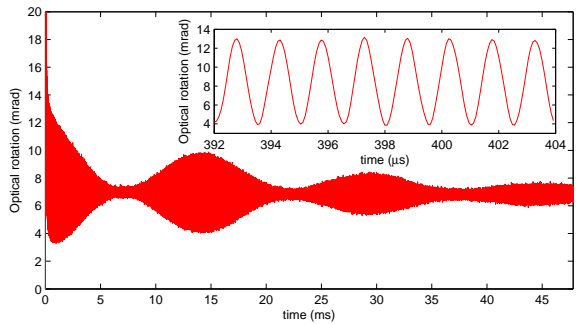}
    \caption{(Color online) An example of data showing optical rotation of the probe
    light. At time = 0 ms, the pumping stage ended.
    The signal undergoes several collapses and revivals during the
    decay. The magnetic field was set to 0.477 G.
    The inset shows a zoom-in revealing the fast oscillation at $2\nu_L=667$ kHz.}
    \label{Fig_RawData}
    \vspace{-.5cm}
\end{figure}

Figure \ref{Fig_Apparatus} shows the experimental setup. A paraffin-coated cylindrical glass cell (with diameter and length of 2 cm) at room temperature containing $^{87}$Rb is located within a multi-layer magnetic shield with a system of coils in the inner volume designed to produce a homogeneous Earth-range magnetic field along the $x$ direction. The atoms in the $F=2$ ground-state are optically pumped and probed using a beam from a diode laser (L1) with frequency close to either the D1 or D2 line.
The intensity of the light is modulated with an acousto-optical modulator (AOM). The same laser beam, initially linearly polarized in the $z$ direction, is used to create atomic alignment (the rank-two atomic polarization that is characterized by a preferred ``alignment axis'') and to probe the alignment via optical rotation due to the polarized atoms \cite{Gaw06}. A measurement consists of two consecutive stages. During the first, pumping stage, the laser intensity is square-wave modulated (duty cycle 1/8) at a rate equal to twice the Larmor frequency, and ground-state atomic alignment is synchronously pumped  for $3$ ms. The beam diameter is $\approx 3$ mm, and the power during the ``on'' part of the cycle is $\approx 5$ mW. Once the atoms are pumped, the AOM is set to transmit around $4\ \mu$W of light, which probes the atomic alignment created during the pumping stage. The optical rotation of the transmitted light is measured by a balanced polarimeter.
An example of a free-induction decay (FID) signal (averaged over 1024 cycles) is shown in Fig.\ \ref{Fig_RawData}. The FID signal oscillates at 667 kHz (see inset) corresponding to twice the Larmor frequency. The NLZ-induced beats in the signal are observed as a change in the oscillation amplitude with time. The signal has an offset of around 7 mrad due to imperfect balancing of the polarimeter.

We model the measured FID signals with a function of the form
\begin{eqnarray}
S(t)&=& \left\{ C_2\cos\prn{2\pi\sbrk{2\nu_L-\delta}t-\alpha}
+ C_1\cos\prn{4\pi\nu_Lt}  \right.\nonumber \\
&&    \left.+ C_2\cos\prn{2\pi\sbrk{2\nu_L+\delta}t+\alpha}\right\}e^{-t/T} , \label{eq:LI_Signal}
\end{eqnarray}
which consists of a central component of amplitude $C_1$ oscillating
at twice the Larmor frequency and two sidebands with frequencies
$2\nu_L \pm \delta$ of equal amplitude $C_2$ and opposite phase $\pm
\alpha$. The three frequency components decay with time, and for simplicity
it is assumed that the $1/e$ decay time $T$ is the same for all three components.
We analyze the envelope of the signal by postprocessing 
the data with a digital lock-in amplifier with reference frequency $2\nu_L$. The envelope $R(t)$ can be calculated from Eq.\ (\ref{eq:LI_Signal}) and we find
\begin{eqnarray}
R\left(t\right) & \approx &  \left[ \frac{1}{4} C_1^2+\frac{1}{2}C_2^2+C_1 C_2 \cos \left(2\pi\delta t-\alpha \right) \right. \nonumber \\
&&\left. +\frac{1}{2}C_2^2 \cos \left(4\pi\delta t-2\alpha \right)\right]^{1/2}\cdot e^{-t/T}. \label{Eq_R}
\end{eqnarray}
It is found experimentally that the demodulated FID signals 
are well described by Eq.\ (\ref{Eq_R}). 
For the data presented in Fig.\ \ref{Fig_RawData}, a fit of the envelope to Eq.\ (\ref{Eq_R}) gives
the value of $\delta=65.37(7)$ Hz, consistent with the expected
value due to the NLZ effect for the magnetic field of $B=0.477$ G
used for the data presented in Fig.\ \ref{Fig_RawData}.
The overall $1/e$ decay time extracted from the fit is $20.8(7)$ ms, limited by
magnetic-field gradients and drifts in the bias magnetic field
combined with signal averaging [for comparison, at lower magnetic
field of 24 mG the relaxation time was measured to be $55(2)$ ms].
\begin{figure}
    \includegraphics[width=3 in]{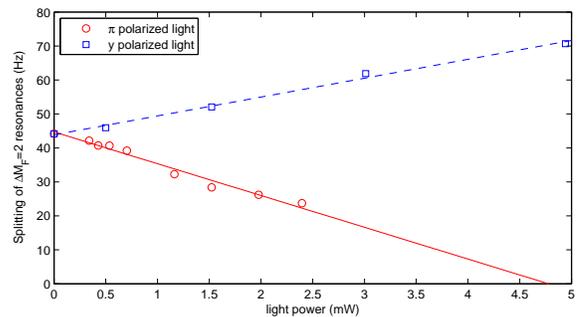}
    \caption{(Color online) Splitting between adjacent magnetic resonances (see Fig.\ \ref{Fig_Fourier}) as a function of L2 light power 
    for polarizations along (circles) and transverse (squares) to the magnetic field (denoted $\pi$ and $y$ polarization
    in the figure). Lines are linear fit to the data. The magnetic field was set to 0.397 G.}
    \label{Fig_Splitting}
\end{figure}
\begin{figure}
    \includegraphics[width=3 in]{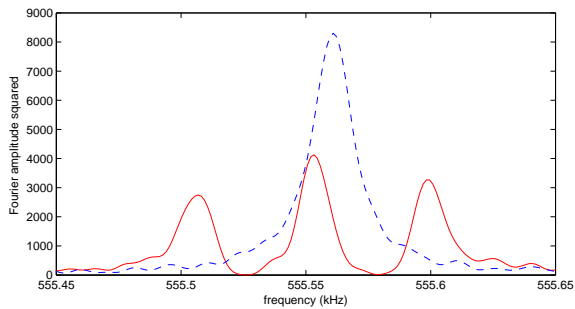}
    \caption{(Color online) Power spectrum of a FID signal
    showing the splitting of the $\Delta M_F=2$ magnetic resonances due to
    NLZ (solid line), and of a signal where NLZ was
    compensated by light shifts induced by the L2 light (dashed line). The magnetic field was set to 0.397 G.}
    \label{Fig_Fourier}
    \vspace{-.5cm}
\end{figure}

In order to cancel NLZ, linearly polarized light from a
second diode laser (L2) tuned close to the D1 resonance
was directed
through the vapor cell along the $z$ direction. 
The splitting of the $\Delta M_F=2$ resonances
was measured as described above for different detunings,
polarizations and powers of the L2 light. Figure \ref{Fig_Splitting}
shows the splitting as a function of light power inside the cell for two
polarizations, along and transverse to the magnetic field (denoted
$\pi$ and $y$ polarization in the figure), for a detuning
$\Delta=-3.8$ GHz from the $F=2 \rightarrow F'=1$ D1 transition. The
splittings were fit to straight lines yielding the slopes $-9.4(13)$ Hz/mW 
and $5.6(8)$ Hz/mW, which within the uncertainties confirms 
that the light shifts due
to $y$ polarized light have the opposite sign and are of half the
magnitude compared to the light shifts due to $\pi$ polarized light. It
should be noted that when the splitting is smaller than the
linewidth of the resonances, it is difficult to extract the
splitting from the FID signal. Therefore, in order to find the light
power where the light shifts cancel NLZ, an extrapolation
from the linear fit was used. In Fig.\ \ref{Fig_Splitting} the light
power where the splitting is zero
is $4.8(7)$ mW.

Figure \ref{Fig_Fourier} (solid line) shows the power spectrum of a FID
signal with the L2 light blocked. Three $\Delta M_F=2$ resonances are seen, 
split from each other in frequency due to NLZ. 
When the L2 light is on (dashed line), with the
appropriate power needed to cancel NLZ, the peaks are combined into a single, stronger resonance.
However, the L2 light also broadens the resonance.
The FWHM of the resonances are 14.0(4) Hz (solid line) and 27.8(4) Hz (dashed line), giving a broadening of 13.8(6) Hz due to L2 light.
By comparing the two spectra in Fig.\
\ref{Fig_Fourier}, it is seen that there is a shift of $\approx 8$ Hz in
the central frequency of the resonances. The shift is not
thought to be due to the L2 light, but instead due to drifts in the
current supply for the bias magnetic field in between the two
measurements.
\begin{figure}
    \includegraphics[width=3 in]{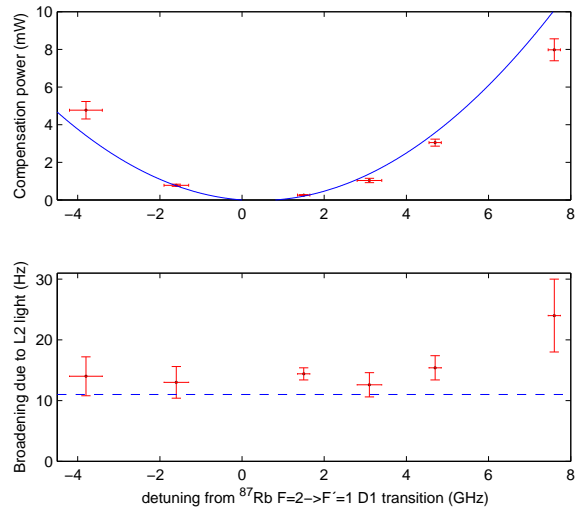}
    \caption{(Color online) Top: L2 light power needed to cancel NLZ at
    a magnetic field of 0.397 G as a function of L2 laser frequency.
    Dots represent measured values, and the solid line is the theoretical 
    calculation given by Eq.\ (\ref{PowerD1Analytical}).
    Bottom: Broadening of the magnetic resonance due to the L2 light at the compensation power.
    Dashed line is the 11 Hz estimate due to dephasing given by Eq.\ (\ref{GammaDephase}).}
    \label{Fig_DeltaSplitting}
    \vspace{-.5cm}
\end{figure}

The L2 light power needed to cancel the NLZ effect was measured for
different frequencies of the L2 light close to the D1 resonance;
the results are plotted in Fig.\ \ref{Fig_DeltaSplitting} (top). For these
measurements, the
magnetic field was set to $B=0.397$ G, producing a NLZ shift 
$\delta \approx 45$  Hz.
Also plotted in Fig.\ \ref{Fig_DeltaSplitting} (top) is the theoretical calculation
for the compensation power given by Eq.\ (\ref{PowerD1Analytical}),
showing a reasonable
agreement between theory and experiment. 
The broadening of the magnetic resonance due to the L2 light at the compensation power was also measured and is 
plotted in Fig.\ \ref{Fig_DeltaSplitting} (bottom)
together with the estimated broadening due to dephasing. In the frequency range within -4 to 5 GHz
of the $F=2 \rightarrow F'=1$ D1 transition the broadening is approximately constant and consistent with the dephasing estimate.
For the detuning $\Delta=7.6$ GHz (which is on the $F=1 \rightarrow F'=2$ resonance) a slightly larger broadening was measured.
We note that the broadening due to dephasing can be avoided if the compensating light has a homegeneous intensity profile over the cell volume. 
In the experiment the beam diameter of the compensating light was rather small, and we therefore expect that it is possible to significantly reduce the broadening by for instance expanding the beam.

\section{Discussion and Conclusions}
To be useful in practical magnetometers, our method for compensating NLZ should be robust against, for instance, intensity fluctuations of the L2 laser light. 
At Earth-range magnetic fields, NLZ splits the magnetic resonances considered in the experiment by approximately 70 Hz. 
We compensate NLZ by overlapping the resonances.
The resonances will overlap if they are positioned within their width, which in our particular case is around 30 Hz (see Fig.\ \ref{Fig_Fourier}). Since the splitting between the resonances is linear in L2 laser intensity, the relative intensity noise of the laser should be (much) less than the ratio of the linewidth to the splitting without compensation, in this case 30 Hz / 70 Hz $\approx$ 40 \%. In practical situations one can easily stabilize lasers to have less than 1 \% intensity noise. The intensity noise has therefore very little effect on the compensation. 
Similarly, we estimate that the frequency drifts of the L2 laser within easily achievable $\pm 1 $MHz lead to sub-1-\% changes in the induced light shifts, much
smaller than typical resonance width.

In conclusion, using AC Stark shifts, we compensated the nonlinear Zeeman effect for $^{87}$Rb atoms located in a magnetic field comparable to the Earth's magnetic field. 
The method can directly be applied to alkali-vapor atomic magnetometers in order to reduce heading errors and increase sensitivity.

\section*{ACKNOWLEDGMENTS} 
The authors are grateful to ~E.~S.~Polzik for encouragement and support, to W. Gawlik for comments on the manuscript and to E. Corsini for helpful discussions.
This work has been supported by NURI grant HM1582-08-1-0006, ONR MURI and STTR grants.

\bibliographystyle{apsrev}
\bibliography{kjensen}

\end{document}